# Phase-multiplexed optical computing: Reconfiguring a multi-task diffractive optical processor using illumination phase diversity

Xiao Wang[1,2,3] and Aydogan Ozcan[1,2,3]

[1]Electrical and Computer Engineering Department, University of California, Los Angeles, CA, 90095, USA
[2]Bioengineering Department, University of California, Los Angeles, CA, 90095, USA
[3]California NanoSystems Institute (CNSI), University of California, Los Angeles, CA, 90095, USA
Corresponding to: ozcan@ucla.edu

## Abstract

Optical computing based on diffractive networks has attracted broad attention due to its low-latency, low energy consumption, and inherent parallelism. Prior work has shown that, given sufficient degrees-of-freedom, a diffractive network composed of spatially optimized surfaces can all-optically implement an arbitrary complex-valued linear transformation between an input and output aperture. Here, we report a monochrome multi-task diffractive network architecture that leverages illumination phase multiplexing to dynamically reconfigure its output function and accurately implement a large group of complex-valued linear transformations between an input and output aperture. Each member of the desired group of $T$ unique transformations is encoded and addressed with a distinct 2D illumination phase profile, termed "phase key", which illuminates the input aperture, activating the corresponding transformation at the output field-of-view. A common diffractive optical network, optimized with $T$ phase keys, demultiplexes these encoded inputs and accurately executes any of the $T$ distinct linear transformations at its output. We demonstrate that a diffractive network composed of $N = 2TN_iN_o$ optimized diffractive features can realize $T$ distinct complex-valued linear transformations – accurately executed for any complex field at the input aperture, where $N_i$ and $N_o$ refer to the input/output pixels, respectively. In our proof-of-concept numerical analysis, $T = 512$ complex-valued transformations are implemented by the same monochrome diffractive network with negligible error using illumination phase diversity. Compared with wavelength-multiplexed diffractive systems, phase-multiplexing architecture significantly lowers the transformation errors, potentially enabling larger-scale optical transformations to be implemented through a monochrome processor. Phase-multiplexed multi-task diffractive networks would enhance the capabilities of optical computing and machine-vision systems.

**Keywords**: Diffractive processors, diffractive networks, phase-multiplexed computing, reconfigurable diffractive processors; optical computing

# Introduction

In this current information era, the explosive growth of data and the escalating demand for information processing have made the efficient execution of large-scale workloads a central driver of computing. As transistor scaling reaches physical limits and on-chip power density continues to increase, conventional electronic digital computing is approaching its boundaries in terms of compute density, energy efficiency, and parallelism. Optical computing[1–7], owing to its low energy consumption, low latency and massive parallelism, has been explored to meet surging compute demands. By processing information using optical waves, computation can be realized through the physics of light propagation itself, potentially offering reductions in energy consumption alongside increased compute density – especially if the information of interest is in the analog domain. A variety of optical-computing architectures have been explored, including on-chip integrated photonic circuits[5,8–10], optical metamaterials[11–15], diffractive deep neural networks ($D^2NN$)[16–23], among others[24–26]. As an emerging free-space-based computing approach, a typical $D^2NN$[16] architecture consists of multiple cascaded transmissive (or reflective) diffractive layers, each comprising tens of thousands of trainable diffractive features. During the training of a $D^2NN$, the transmission/reflection coefficients of these features are optimized for a desired computational task via error backpropagation. Through light–matter interactions, the trained layers modulate the phase and amplitude of the incident field to all-optically perform a desired computational task between an input and output aperture. Importantly, these diffractive layers are passive optical components that require no input power other than the illumination source. Consequently, a $D^2NN$ performs its target computation in a single optical pass through a thin, passive optical volume, delivering parallelism at the diffraction limit of light with low latency and negligible energy consumption. $D^2NN$s have been demonstrated for diverse information-processing tasks, including, e.g., object classification[16,17,27–31], nonlinear function approximation[32,33], quantitative phase imaging[34–36], super-resolution image displays[37], information encryption[38–40], logical operations[41–43] and universal linear transformations.[18–23]

Here, we report a phase-multiplexed multi-task diffractive optical processor that reconfigures its output function using illumination phase diversity to implement a large set of arbitrarily selected complex-valued linear transformations between an input and output aperture. In this design, the incident optical wave has specific phase patterns, termed 'phase keys', that serve as multiplexing channels illuminating the input aperture of a monochrome diffractive processor (**Fig. 1**). These phase keys are trainable for a given set of desired transformations/tasks and are jointly optimized with a single demultiplexing $D^2NN$. Once trained, the fixed monochrome $D^2NN$ can reconfigure its output function using different illumination phase keys and universally execute any of the $T$ pre-programmed transformations simply by switching the phase key at the illumination plane from one optimized pattern to another; see **Fig. 1**. We show that, when the number ($N$) of trainable diffractive features of the $D^2NN$ architecture satisfies $N \geq 2TN_iN_o$, the approximation error of each complex linear transformation/function becomes negligible. Numerically, we demonstrate $T = 512$ unique transformations; considering acceptable error bounds for different applications, we also show that the number of linear transformations executed through the same monochrome $D^2NN$ can be extended to e.g., $T > 2,000$. This large-scale multiplexed information processing capability of phase-multiplexed diffractive processors can be further enhanced by increasing $N$ or by integrating additional multiplexing modalities such as polarization[19,44] and wavelength encoding[18], thereby improving the overall system throughput. This reconfigurable multi-tasking framework is broadly applicable across different parts of the electromagnetic spectrum, including the visible band. We anticipate that the presented phase-multiplexed multi-tasking diffractive processors will accelerate the development of optical information processing and machine vision systems.

# Results

Throughout this work, the terms "diffractive deep neural network", "diffractive neural network", "diffractive optical network", and "diffractive processor/network" are used interchangeably to refer to the same class of free-space optical computing architectures. **Fig. 1** illustrates the framework of our phase-multiplexed optical processor, which consists of cascaded diffractive layers comprising a total of $N$ diffractive features. These $N$ features are uniformly distributed across the layers, and their phase profiles are treated as trainable parameters optimized through stochastic gradient descent and error backpropagation to perform arbitrarily-selected linear transformations between the input and output apertures. Note that the eight-layer configuration shown in **Fig. 1** serves as an illustrative example; the layer count can be tailored according to the total number ($N$) of diffractive features desired, which will be detailed below.

The objective of this work is to enable a single monochrome diffractive network to reconfigure its output function using illumination phase diversity and accurately perform a large number of distinct linear transformations for an infinite number of different complex input fields without requiring the re-training or re-fabrication of the diffractive layers for each transformation. To achieve this goal, we adopt an illumination phase-multiplexing strategy. Specifically, a set of $T$ phase-encoded beams sequentially illuminate the input field and are demultiplexed by a single, common diffractive processor to execute $T$ different linear transformations at the output aperture. Accordingly, we assign a distinct encoding phase ('phase key') for each of the $T$ target transformations, with these keys collectively establishing $T$ different illumination phase channels used for multiplexing. All $T$ phase keys are treated as trainable parameters and jointly optimized with the $N$ diffractive features during training. After training, both the set of $T$ phase keys and the diffractive layers are *fixed*.

To evaluate the diffractive processor's ability to perform large-scale linear transformations, we construct $T$ independent random complex-valued matrices ($A_1, A_2, \ldots, A_T$) each with dimensions $N_i \times N_o$, where $N_i$ and $N_o$ refer to the number of input and output pixels that make up the input and output apertures, respectively. To confirm the statistical independence among these matrices, the cosine similarity between all possible matrix pairs is also computed (see Supplementary **Fig. S1**). These complex matrices serve as target linear transformations (functions) between the input and output fields, expressed as $o_{A_t,n} = A_t i_{A_t,n}$, where $i_{A_t,n}$ and $o_{A_t,n}$ denote the vectorized complex-valued input and output fields for the $t^{\text{th}}$ channel, respectively. For each phase channel, the complex input field $i_{A_t,n}$ (to be accurately transformed) can be arbitrarily chosen and the processor supports an infinite number of input fields/samples, i.e., $n = 1, 2, \ldots, \infty$, for any phase channel. For a given channel $t$, the illumination beam is modulated by the phase key $\varphi_{A_t}$ before interacting with the input field to be transformed. The resulting wavefront propagates through the static, optimized diffractive system, yielding the output field $o'_{A_t,n} = \text{D}^2\text{NN}\{e^{j\varphi_{A_t}} \cdot i_{A_t,n}\}$, where $\text{D}^2\text{NN}\{\cdot\}$ denotes the diffractive optical network operator. Consequently, the convergence between the diffractive output field $o'_{A_t,n}$ and the theoretical ground truth field $o_{A_t,n}$ for any arbitrarily selected complex input field indicates that the implemented optical transformation, $A'_t$, serves as a high-fidelity approximation of the target linear transformation $A_t$, i.e., $A_t \approx A'_t$ for each channel $t$.

To train the diffractive neural network to perform the aforementioned multiplexed linear transformations, a total of 70,000 complex-valued input field vectors $i_n$ were randomly generated. For each target linear transformation matrix $A_t$, the corresponding ground truth output field vectors were computed as $o_{A_t,n} = A_t i_{A_t,n}$. In other words, each target transformation matrix was associated with 70,000 paired input–output complex field examples. These sample pairs were divided into a non-overlapping training set, validation set, and test set containing 55,000, 5,000, and 10,000 samples, respectively. Without loss of generality, the operating wavelength of the all-optical diffractive processor is set to $\lambda = 0.8$ mm. We note that the proposed framework is wavelength-agnostic and can be broadly adapted across the electromagnetic spectrum by scaling the optimized features proportional to the illumination wavelength. Light propagation within the system is modeled using the scalar diffraction theory as detailed in the **Materials and Methods** section. To ensure accuracy in our forward model, both the simulation sampling period and the lateral size of the individual diffractive features are set to $0.5\ \lambda$. In our analyses, the sizes of the input and output fields are set to $N_i = N_o = 5 \times 5 = 25$. To compare the performances of different multiplexing-channel configurations, we maintained the same lateral size for the diffractive layers, thereby preserving a consistent input/output numerical aperture (NA) across all the configurations being compared. Consequently, the total number of trainable diffractive features ($N$) is adjusted by varying the layer count without impacting the input/output NA.

To evaluate the performance of the diffractive processor in implementing phase-multiplexed linear transformations, we first select $T = 4, 8, 16,$ and $32$ phase channels. The exploration of larger-scale multiplexing and potential upper limits in performance is presented in the later analysis shown below. For the number of trainable diffractive features $N$, we chose $N \approx 2TN_iN_o$. Thus, for $T = 4, 8, 16,$ and $32$, we used $N = 5{,}041, 10{,}082, 20{,}164,$ and $40{,}328$, respectively. Given that each diffractive layer has a fixed size to preserve NA, these phase-multiplexed configurations correspond to optical systems comprising 1, 2, 4, and 8 cascaded diffractive layers, respectively.

During the training of a diffractive processor with a given $T$, both the phase keys and the diffractive network layers were jointly optimized by minimizing the mean square error (MSE) between the diffractive output and the target output (ground truth) fields. Taking $T = 8$ phase channels as an example, **Fig. 2** shows the learned phase keys (col. 1), the corresponding target linear transformation matrices $\boldsymbol{A}_t$ (cols. 2-3), the all-optical linear transformation matrices $\boldsymbol{A}'_t$ (cols. 4-5), and their absolute error (col. 6). A comparison between the amplitude and phase distributions of the all-optical transformations and their target counterparts, together with the absolute error maps, clearly reveals that the all-optical complex linear transformations achieved through phase-multiplexing precisely match their ground-truth targets. Furthermore, **Fig. 3** presents some examples of complex-valued input and output fields for the $T = 8$ configuration. It can be observed that the diffractive processor accurately performs all-optical linear transformations in every phase key channel. For randomly selected test input fields (never seen before) across the various phase channels, the corresponding output complex fields exhibit an excellent match with their ground-truth targets, with negligible error in both the amplitude and phase channels. These results are consistent with the transformation matrix comparisons reported in **Fig. 2**.

Next, we analyze the linear transformation performance of the individual phase channels in our multiplexed diffractive processors. **Fig. 4** presents the channel-wise transformation error $\mathrm{MSE_{Transformation}}$ and output-field error $\mathrm{MSE_{output}}$ for the phase-multiplexed diffractive networks with $T = 4, 8, 16,$ and $32$. The results confirm that the monochrome diffractive processor maintains negligible error across all phase key channels, with a strong correlation observed between $\mathrm{MSE_{Transformation}}$ and $\mathrm{MSE_{output}}$. Although minor variations in error are observed among different channels, the overall performance remains comparable, indicating that the diffractive processor does not exhibit a bias toward specific phase-multiplexed channels or target transformations.

The preceding results demonstrated that, combined with illumination phase encoding, a single monochrome diffractive network can perform a set of $T$ arbitrarily-selected complex-valued linear transformations with negligible error. However, as the number of multiplexed channels increases, both the linear transformation error and the output-field error exhibit a gradual rise. To investigate the interplay between diffractive processor performance and multiplexing capability, and to determine the potential limits of achievable multiplexing, we extend our analysis to include a larger number of phase-multiplexed channels. Specifically, we maintain $N_i = N_o = 25$ and select $T = 32, 64, 128, 256,$ and $512$. For each phase multiplexing level $T$, two diffractive processor models were trained with different numbers of diffractive features, specifically $N = 2TN_iN_o$ and $N = 3TN_iN_o$. A total of 512 randomly generated complex-valued matrices of $25 \times 25$ were used as target linear transformation matrices. The cosine similarity values among these matrices are close to zero, as shown in Supplementary **Fig. S1**. It is important to emphasize that the maximum number of multiplexed channels considered here, $T = 512$, does not represent the upper limit of the proposed phase-multiplexed diffractive processor, but rather reflects a practical constraint imposed by our available academic computational resources.

**Fig. 5** further quantifies the impact of the channel multiplexing scale $T$ on the diffractive processor's all-optical linear transformation performance, depicting the transformation error ($MSE_{Transformation}$, **Fig. 5a**) and the output-field error ($MSE_{output}$, **Fig. 5b**). As expected, both error metrics increase with the number of multiplexed channels $T$, yet remain at remarkably low levels. Moreover, the model with an increased number of diffractive features $N = 3TN_iN_o$ consistently outperforms the baseline model $N = 2TN_iN_o$, exhibiting improved optical transformation fidelity. As an example, for $T = 512$, the model with $N = 2TN_iN_o$ achieves a transformation error of $MSE_{Transformation} = 3.86 \times 10^{-8}$ and an output-field error of $MSE_{output} = 4.58 \times 10^{-9}$. In comparison, the corresponding errors for the model with $N = 3TN_iN_o$ are $MSE_{Transformation} = 2.19 \times 10^{-8}$ and $MSE_{output} = 2.62 \times 10^{-9}$, respectively.

Next, we extrapolate the behavior of the diffractive processor shown in **Fig. 5a** and **5b** to multiplexing scales beyond $T = 512$ to estimate the level of error for larger values of $T$. The fitted curves (dashed lines) indicate that even for a multiplexing scale of $T = 10,000$, the transformation error $MSE_{Transformation}$ could remain within a range of $10^{-6} - 10^{-4}$. This suggests that $T = 10,000$ is far from representing the multiplexing limit of the proposed diffractive processor. However, due to computational constraints, our numerical implementation has been limited to $T = 512$. To obtain a more conservative estimate, we set error thresholds of $MSE_{Transformation} = 9 \times 10^{-7}$ and $MSE_{output} = 10^{-7}$ (see the gray dashed lines). Under these transformation error bounds, the model with $N = 3TN_iN_o$ diffractive features is projected to implement up to approximately $T \sim 2400$ distinct linear transformations multiplexed using different phase keys (orange dashed line). For the model with $N = 2TN_iN_o$, the same analysis yields an upper bound of $T \sim 1800$ transformations (green dashed line). It is noteworthy that in earlier work on wavelength-multiplexed broadband diffractive optical processors[18], the system scaled to $\sim 2000$ linear transformations with error thresholds of $MSE_{Transformation} = 9 \times 10^{-3}$ and $MSE_{output} = 10^{-3}$. In contrast, at a comparable scale of $\sim 2400$ transformations, the phase-multiplexed monochrome diffractive processor achieves error levels that are orders of magnitude lower. This highlights the superior scalability of the proposed phase-multiplexing-based monochrome approach, underscoring its significance for large-scale all-optical computing.

Next, we analyze the layer bit depth of a diffractive optical network, a critical parameter that significantly influences its optical approximation performance and is closely related to the practical fabrication of diffractive layers. In practice, for the fabrication of each diffractive layer, a minimum thickness is required as a base substrate for the optimized layer. Consequently, the bit depth is determined by the allowable quantization levels between this baseline and the maximum thickness of the optimized layer. For example, a diffractive layer with an 8-bit resolution has $2^8 = 256$ distinct thickness levels for each diffractive feature. Here, we consider $T = 8$ phase-multiplexed transformation channels and analyze two diffractive processor designs with $N = 2TN_iN_o$ and $N = 3TN_iN_o$. Each design is trained under different bit depth requirements (2, 4, 8, 12, and 16 bits) to examine the impact of layer bit-depth quantization on the all-optical linear transformation performance. **Fig. 6** summarizes the performance metrics for various diffractive designs constrained with different bit depths. It is evident that when the bit depth is $\geq 8$, the transformation error of the diffractive processor becomes negligible. Conversely, coarse quantization (e.g., 2-bit depth) yields a sharp rise in the all-optical transformation errors. Nevertheless, a comparison between the $N = 2TN_iN_o$ and $N = 3TN_iN_o$ diffractive models reveals that increasing the number of diffractive features can partially mitigate the degradation in performance caused by a significantly reduced bit depth.

The output diffraction efficiency ($\eta$) is another critical factor for diffractive optical networks, as it directly affects the practical applicability of this framework. The factors influencing diffraction efficiency can be categorized into two main groups. The first arises from the intrinsic properties of the diffractive material: when light interacts with the diffractive material, part of the optical energy is inevitably absorbed and reflected. The second factor comes from energy losses due to light leakage out of the physical aperture of the diffractive processor during the forward propagation. The latter can be mitigated by introducing a penalty term[18,21] related to diffraction efficiency into the training loss function (see **Materials and Methods** for details). Taking $T = 8$ as an example, we consider two designs of diffractive processors with $N = 2TN_iN_o$ and $N = 3TN_iN_o$. During training, we impose diffraction efficiency-related penalty thresholds of $\eta_{th} = 1\%, 5\%, 10\%, 15\%$ and $20\%$. The results of these diffractive network designs are shown in **Fig. 7**, where the solid curves represent the transformation errors as a function of the penalty threshold, and the dashed curves show the transformation errors as a function of the actual realized output diffraction efficiency achieved by the resulting diffractive network. As the penalty threshold used in training increases, both the error of the all-optical transformation and the realized output diffraction efficiency increase, as expected. For $N = 2TN_iN_o$, when the penalty threshold is set to 20%, the optimized diffractive optical processor achieves transformation errors of $\text{MSE}_{\text{Transformation}} = 1.12 \times 10^{-6}$ and $\text{MSE}_{\text{output}} = 1.37 \times 10^{-7}$, while the actual output diffraction efficiency reaches ~23.32%. It is worth emphasizing that an error on the order of $10^{-7} - 10^{-6}$ remains negligible and can be further reduced by increasing the number of optimizable diffractive features, $N$.

## Discussion

We presented a phase-multiplexed diffractive processor capable of performing large-scale, arbitrary, complex-valued linear transformations using a single monochrome diffractive network that reconfigures its output function using illumination phase diversity. These phase multiplexed transformations are executed optically by passive diffractive layers, requiring no energy beyond the illumination light. We further investigated and quantified the potential limits of the multiplexing channel number $T$, the bit depth of the diffractive network, and the output diffraction efficiency of these all-optical transformations.

Previous research demonstrated polarization multiplexing[19] using a static array of polarizers to realize four independent linear transformations, rotational-angle multiplexing[23] of diffractive layers to implement $\sim 200$ permutation operations, and wavelength multiplexing[18] to simultaneously achieve >180 arbitrary complex-valued linear transformations, with scalability up to $\sim 2000$ unique transformations. In contrast, the phase-multiplexed monochrome diffractive processor reported here achieves 512 arbitrary complex-valued linear transformations and can be scaled up to $\sim 2400$ transformations with a negligible increase in error. The phase-multiplexing-based monochrome diffractive processors achieve lower transformation errors, demonstrating a four orders-of-magnitude improvement over wavelength-multiplexing-based broadband approaches. This highlights the substantial potential of phase multiplexing for further enhancing both computational throughput and precision in all-optical computing systems.

It is important to emphasize that an alternative, parallelized strategy that engineers $T$ independently optimized diffractive networks, each using $\sim N/T$ diffractive features, cannot serve as an effective replacement for the phase-multiplexing scheme reported here. Independent diffractive networks require their own optical routing paths, and the resulting large-scale optical interconnects would introduce additional noise, losses, and alignment errors, thereby degrading both energy efficiency and computational accuracy. Moreover, switching between multiple independent diffractive networks would significantly increase system complexity. As the multiplexing scale expands to larger values, such as $T \geq 512$, this issue would become even more challenging to handle; on the other hand, our phase-multiplexed multi-tasking diffractive architecture is capable of executing a large group of transformations within a single monochrome system.

It is worth noting that the proposed phase-multiplexed diffractive processor also has its limitations. For example, the form of phase multiplexing imposed on the illumination field inherently prevents large-scale *parallel* execution of multiple transformations. Similar to wavelength multiplexing, phase multiplexing also utilizes a single diffractive neural network to perform large-scale linear transformations; however, it must operate sequentially, meaning that only one target transformation can be performed at a time. Furthermore, because each linear transformation is assigned to a different phase key, additional optical components or phase modulators are required to customize the phase profile of the incident illumination required for multiplexing. This inevitably increases the overall system complexity. Nevertheless, these limitations do not outweigh the advantages of the phase-multiplexing approach in terms of transformation scale and performance. Moreover, it is foreseeable that combining phase multiplexing with other physical degrees of freedom, such as polarization and wavelength diversity, could further enhance the multiplexing capacity while alleviating some of these limitations.

Several other factors, such as lateral and axial misalignments of the diffractive layers, surface reflections, and fabrication-induced imperfections, can also limit the performance of a diffractive optical processor. These issues can be potentially mitigated through advanced fabrication techniques, including high-precision lithography and the use of anti-reflective coatings. Furthermore, a "vaccination" strategy can be adopted during training, in which some of these imperfections and random misalignments are incorporated as stochastic perturbations in the forward model to alleviate their impact on testing performance. Previous experimental demonstrations[18,23,38,45] have shown that such practical, hard-to-control errors and imperfections do not lead to significant discrepancies between the experimental results and numerical simulations, demonstrating the success of these vaccination strategies.

During the training of phase-multiplexed diffractive neural networks, the phase keys and the diffractive layers are jointly optimized through stochastic gradient descent and error backpropagation. Specifically, the loss is propagated through the diffractive neural network and distributed in parallel to every illumination phase channel. The losses from all the channels are then summed to form the total loss, which is subsequently used to update all the learnable parameters (as detailed in the **Materials and Methods**). Using the total loss to update/optimize the diffractive layer features is practical, since all the transformation channels share the same set of diffractive layers in their forward model. However, the phase keys associated with different transformation channels are inherently meant to be used independent of each other, i.e., one phase key for each transformation function. Using the total loss to jointly update all the phase keys introduces some cross-talk, thereby reducing the independence of the phase keys. To quantify this effect, we computed the pairwise cosine similarity among the 32 optimized phase keys obtained from a model with $T = 32$ and $N = 2TN_iN_o$ (see Supplementary **Fig. S2**). The results show that the cosine similarity values are ~0.85, indicating a relatively strong correlation among the phase keys. We attribute this correlation to the use of a total-loss–based optimization strategy. A sequential and independent loss-propagation scheme, in which each transformation channel optimizes its own phase key separately, could potentially reduce the correlation among phase keys, thereby mitigating inter-channel cross-talk and achieving superior all-optical transformation performance. However, adopting such sequential optimization would substantially increase the computational burden, which becomes harder when the number of multiplexed channels $T$ is large. Exploring this class of optimization strategies is therefore beyond the scope of the present study and is left as future work.

Finally, we emphasize that phase-multiplexed reconfigurable monochrome diffractive processors exhibit substantial potential in both the fidelity and scalability of large-scale all-optical complex-valued transformations, which is of significant importance for improving the throughput of multiplexed optical computing systems. The presented framework can be extended to other regions of the electromagnetic spectrum, including the visible and IR, by scaling the optimized diffractive features proportional to the illumination wavelength.

## Materials and Methods

### Forward Model of Diffractive Networks

A diffractive optical network comprises a series of diffractive layers. These layers are modeled as thin dielectric optical modulation elements with varying thicknesses. For the $m^{\text{th}}$ feature located at spatial coordinate $(x_m, y_m, z_m)$ on the $k^{\text{th}}$ layer, the complex-valued transmission coefficient $t^k$ is expressed as

$$t^k(x_m, y_m, z_m) = a^k(x_m, y_m, z_m)\exp\left(j\phi^k(x_m, y_m, z_m)\right), \tag{1}$$

where $a$ and $\phi$ are the amplitude and phase coefficients, respectively. The diffractive layers are connected to each other by free-space propagation, which is modeled through the Rayleigh-Sommerfeld diffraction equation:

$$f_m^k(x, y, z) = \frac{z - z_i}{r^2}\left(\frac{1}{2\pi r} + \frac{1}{j\lambda}\right)\exp\left(\frac{j2\pi r}{\lambda}\right), \tag{2}$$

where $r = \sqrt{(x - x_m)^2 + (y - y_m)^2 + (z - z_m)^2}$ and $j = \sqrt{-1}$. $f_m^k(x, y, z)$ represents the complex-valued field at a spatial location $(x, y, z)$ at a wavelength of $\lambda$, which can be viewed as a secondary wave generated from the source at $(x_m, y_m, z_m)$. As a result, the optical field at location $(x_m, y_m, z_m)$ modulated by the $k^{\text{th}}$ layer ($k \geq 1$, treating the input object plane as the $0^{\text{th}}$ layer) can be written as:

$$E^k(x_m, y_m, z_m) = t^k(x_m, y_m, z_m) \cdot \sum_{n \in S} E^{k-1}(x_n, y_n, z_n) \cdot f_n^{k-1}(x_m, y_m, z_m), \tag{3}$$

where $S$ denotes all the diffractive features on the previous layer.

The amplitude and phase components of the complex transmittance of the $m^{\text{th}}$ feature of diffractive layer k, i.e., $a^k(x_m, y_m, z_m)$ and $\phi^k(x_m, y_m, z_m)$ in **Eq. (1)**, are defined as a function of the material thickness $h_m^k$, as follows:

$$a^k(x_m, y_m, z_m) = \exp\left(-\frac{2\pi\kappa_d(\lambda)h_m^k}{\lambda}\right) \tag{4}$$

$$\phi^k(x_m, y_m, z_m) = (n_d(\lambda) - n_{air})\frac{2\pi h_m^k}{\lambda} \tag{5}$$

where the wavelength-dependent dispersion parameters $n_d(\lambda)$ and $\kappa_d(\lambda)$ represent the refractive index (real part of the complex refractive index) and the extinction coefficient (imaginary part) of the diffractive layer material, respectively: $\tilde{n}_d(\lambda) = n_d(\lambda) + j\kappa_d(\lambda)$. The $n_d(\lambda)$ and $\kappa_d(\lambda)$ curves of the diffractive layer material used for training the diffractive models reported in this paper are shown in **Fig. S3**. The diffractive feature thickness $h_m^k$ is composed of two parts: $h = h_{\text{base}} + h_{\text{trainable}}$. Here, $h_{\text{trainable}}$ represents the learnable thickness parameter for each diffractive feature, with a value range limited between 0 and $h_{\text{max}} = 1.25\lambda$. The additional base thickness $h_{\text{base}}$ is a constant, selected as $0.25\lambda$, which serves as the supporting substrate for the diffractive layer. To achieve the constraint applied to $h_{\text{trainable}}$, an associated latent trainable $h_v$ was defined using: $h_{\text{trainable}} = \frac{h_{max}}{2} \cdot (\sin(h_v) + 1)$. Note that before the training starts, $h_v$ values of all the diffractive features were randomly initialized with a normal distribution (a mean value of 0 and a standard deviation of 1).

## Quantitative metrics for evaluating the performance of all-optical transformations

To quantitatively evaluate the transformation performance of a phase-multiplexed diffractive network, we computed four different metrics for each transformation channel using a blind test dataset: (1) the normalized transformation MSE ($\text{MSE}_{\text{Transformation}}$), (2) the cosine similarity between the all-optical transformation and the target transformation ($\text{CosSim}_{\text{Transformation}}$), (3) the normalized MSE between the output field of the diffractive network and its ground truth ($\text{MSE}_{\text{Output}}$), and (4) the cosine similarity between the output field and its ground truth ($\text{CosSim}_{\text{Output}}$). The transformation error $\text{MSE}_{\text{Transformation,t}}$ of the $t^{th}$ transformation channel is defined as

$$\begin{aligned}\text{MSE}_{\text{Transformation,t}} &= \frac{1}{N_i N_o}\sum_{v=1}^{N_i N_o}|a_t[v] - \mu_t a'_t[v]|^2 \\ &= \frac{1}{N_i N_o}\sum_{v=1}^{N_i N_o}|a_t[v] - \hat{a}'_t[v]|^2,\end{aligned} \tag{6}$$

where $a_t$ is the vectorized form of the ground-truth (target) transformation matrix $\boldsymbol{A_t}$, i.e., $a_t = \text{vec}(\boldsymbol{A_t})$. Likewise, $a'_t$ is the vectorized form of $\boldsymbol{A'_t}$, where $\boldsymbol{A'_t}$ denotes the all-optical transformation matrix implemented by the trained diffractive network. The scalar coefficient $\mu_t$ is introduced to eliminate the scaling mismatch between $A_t$ and $A'_t$ caused by optical losses, and is defined as:

$$\mu_t = \frac{\sum_{v=1}^{N_i N_o} a_t[v]\, {a'}^*_t[v]}{\sum_{v=1}^{N_i N_o} |a'_t[v]|^2} \tag{7}$$

The cosine similarity between the all-optical diffractive transformation and its target (ground truth) for the $t^{\text{th}}$ transformation channel $\text{CosSim}_{\text{Transformation},t}$ is defined as

$$\text{CosSim}_{\text{Transformation},t} = \frac{|a_t^H \hat{a}_t{'}|}{\sqrt{\sum_{v=1}^{N_i N_o} |a_t[v]|^2} \sqrt{\sum_{v=1}^{N_i N_o} |\hat{a}'_t[v]|^2}} \tag{8}$$

The definitions of the normalized MSE between the diffractive network output field and its ground truth for the $t^{\text{th}}$ transformation channel, $\text{MSE}_{\text{Output},t}$, as well as the corresponding cosine similarity $\text{CosSim}_{\text{Output},t}$, follow similar definitions.

**Training Details**

All diffractive optical networks used in this study were trained using PyTorch (v2.3.1, Meta Platforms). We employed the AdamW optimizer with its default hyperparameters for all models. The batch size was set to 16. The learning rate started at 0.001 and decayed by a factor of 0.5 every 10 epochs. Each diffractive network model was trained for 50 epochs, and the best-performing model was selected based on the MSE loss computed on the validation dataset. Model training was accelerated using an NVIDIA H200 GPU. For example, training a phase-multiplexed diffractive network with $T = 512$ and $N = 2TN_i N_o$ required ~66 hours.

**Supporting Information**. The Supporting Information provides:

- Supplementary **Figures S1-S3**
- Training Loss Function

# References

(1) Solli, D. R.; Jalali, B. Analog Optical Computing. *Nature Photon* **2015**, *9* (11), 704–706. https://doi.org/10.1038/nphoton.2015.208.
(2) Athale, R.; Psaltis, D. Optical Computing: Past and Future. *Optics & Photonics News, OPN* **2016**, *27* (6), 32–39. https://doi.org/10.1364/OPN.27.6.000032.
(3) Wetzstein, G.; Ozcan, A.; Gigan, S.; Fan, S.; Englund, D.; Soljačić, M.; Denz, C.; Miller, D. A. B.; Psaltis, D. Inference in Artificial Intelligence with Deep Optics and Photonics. *Nature* **2020**, *588* (7836), 39–47. https://doi.org/10.1038/s41586-020-2973-6.
(4) Shastri, B. J.; Tait, A. N.; Ferreira de Lima, T.; Pernice, W. H. P.; Bhaskaran, H.; Wright, C. D.; Prucnal, P. R. Photonics for Artificial Intelligence and Neuromorphic Computing. *Nat. Photonics* **2021**, *15* (2), 102–114. https://doi.org/10.1038/s41566-020-00754-y.
(5) Feldmann, J.; Youngblood, N.; Karpov, M.; Gehring, H.; Li, X.; Stappers, M.; Le Gallo, M.; Fu, X.; Lukashchuk, A.; Raja, A. S.; Liu, J.; Wright, C. D.; Sebastian, A.; Kippenberg, T. J.; Pernice, W. H. P.; Bhaskaran, H. Parallel Convolutional Processing Using an Integrated Photonic Tensor Core. *Nature* **2021**, *589* (7840), 52–58. https://doi.org/10.1038/s41586-020-03070-1.

(6) Carolan, J.; Harrold, C.; Sparrow, C.; Martín-López, E.; Russell, N. J.; Silverstone, J. W.; Shadbolt, P. J.; Matsuda, N.; Oguma, M.; Itoh, M.; Marshall, G. D.; Thompson, M. G.; Matthews, J. C. F.; Hashimoto, T.; O'Brien, J. L.; Laing, A. Universal Linear Optics. *Science* **2015**, *349* (6249), 711–716. https://doi.org/10.1126/science.aab3642.

(7) Mengu, D.; Rahman, M. S. S.; Luo, Y.; Li, J.; Kulce, O.; Ozcan, A. At the Intersection of Optics and Deep Learning: Statistical Inference, Computing, and Inverse Design. *Adv. Opt. Photon., AOP* **2022**, *14* (2), 209–290. https://doi.org/10.1364/AOP.450345.

(8) Vandoorne, K.; Mechet, P.; Van Vaerenbergh, T.; Fiers, M.; Morthier, G.; Verstraeten, D.; Schrauwen, B.; Dambre, J.; Bienstman, P. Experimental Demonstration of Reservoir Computing on a Silicon Photonics Chip. *Nat Commun* **2014**, *5* (1), 3541. https://doi.org/10.1038/ncomms4541.

(9) Shen, Y.; Harris, N. C.; Skirlo, S.; Prabhu, M.; Baehr-Jones, T.; Hochberg, M.; Sun, X.; Zhao, S.; Larochelle, H.; Englund, D.; Soljačić, M. Deep Learning with Coherent Nanophotonic Circuits. *Nature Photon* **2017**, *11* (7), 441–446. https://doi.org/10.1038/nphoton.2017.93.

(10) Zhang, H.; Gu, M.; Jiang, X. D.; Thompson, J.; Cai, H.; Paesani, S.; Santagati, R.; Laing, A.; Zhang, Y.; Yung, M. H.; Shi, Y. Z.; Muhammad, F. K.; Lo, G. Q.; Luo, X. S.; Dong, B.; Kwong, D. L.; Kwek, L. C.; Liu, A. Q. An Optical Neural Chip for Implementing Complex-Valued Neural Network. *Nat Commun* **2021**, *12* (1), 457. https://doi.org/10.1038/s41467-020-20719-7.

(11) Zangeneh-Nejad, F.; Sounas, D. L.; Alù, A.; Fleury, R. Analogue Computing with Metamaterials. *Nat Rev Mater* **2021**, *6* (3), 207–225. https://doi.org/10.1038/s41578-020-00243-2.

(12) Kwon, H. Nonlocal Metasurfaces for Optical Signal Processing. *Phys. Rev. Lett.* **2018**, *121* (17). https://doi.org/10.1103/PhysRevLett.121.173004.

(13) Ma, W.; Cheng, F.; Liu, Y. Deep-Learning-Enabled On-Demand Design of Chiral Metamaterials. *ACS Nano* **2018**, *12* (6), 6326–6334. https://doi.org/10.1021/acsnano.8b03569.

(14) Mohammadi Estakhri, N.; Edwards, B.; Engheta, N. Inverse-Designed Metastructures That Solve Equations. *Science* **2019**, *363* (6433), 1333–1338. https://doi.org/10.1126/science.aaw2498.

(15) Silva, A.; Monticone, F.; Castaldi, G.; Galdi, V.; Alù, A.; Engheta, N. Performing Mathematical Operations with Metamaterials. *Science* **2014**, *343* (6167), 160–163. https://doi.org/10.1126/science.1242818.

(16) Lin, X.; Rivenson, Y.; Yardimci, N. T.; Veli, M.; Luo, Y.; Jarrahi, M.; Ozcan, A. All-Optical Machine Learning Using Diffractive Deep Neural Networks. *Science* **2018**, *361* (6406), 1004–1008. https://doi.org/10.1126/science.aat8084.

(17) Yan, T. Fourier-Space Diffractive Deep Neural Network. *Phys. Rev. Lett.* **2019**, *123* (2). https://doi.org/10.1103/PhysRevLett.123.023901.

(18) Li, J.; Gan, T.; Bai, B.; Luo, Y.; Jarrahi, M.; Ozcan, A. Massively Parallel Universal Linear Transformations Using a Wavelength-Multiplexed Diffractive Optical Network. *AP* **2023**, *5* (1), 016003. https://doi.org/10.1117/1.AP.5.1.016003.

(19) Li, J.; Hung, Y.-C.; Kulce, O.; Mengu, D.; Ozcan, A. Polarization Multiplexed Diffractive Computing: All-Optical Implementation of a Group of Linear Transformations through a Polarization-Encoded Diffractive Network. *Light Sci Appl* **2022**, *11* (1), 153. https://doi.org/10.1038/s41377-022-00849-x.

(20) Rahman, M. S. S.; Yang, X.; Li, J.; Bai, B.; Ozcan, A. Universal Linear Intensity Transformations Using Spatially Incoherent Diffractive Processors. *Light Sci Appl* **2023**, *12* (1), 195. https://doi.org/10.1038/s41377-023-01234-y.

(21) Kulce, O.; Mengu, D.; Rivenson, Y.; Ozcan, A. All-Optical Synthesis of an Arbitrary Linear Transformation Using Diffractive Surfaces. *Light Sci Appl* **2021**, *10* (1), 196. https://doi.org/10.1038/s41377-021-00623-5.

(22) Yang, X.; Rahman, M. S. S.; Bai, B.; Li, J.; Ozcan, A. Complex-Valued Universal Linear Transformations and Image Encryption Using Spatially Incoherent Diffractive Networks. *APN* **2024**, *3* (1), 016010. https://doi.org/10.1117/1.APN.3.1.016010.

(23) Ma, G.; Yang, X.; Bai, B.; Li, J.; Li, Y.; Gan, T.; Shen, C.-Y.; Zhang, Y.; Li, Y.; Işıl, Ç.; Jarrahi, M.; Ozcan, A. Multiplexed All-Optical Permutation Operations Using a Reconfigurable Diffractive Optical Network. *Laser & Photonics Reviews* **2024**, *18* (11), 2400238. https://doi.org/10.1002/lpor.202400238.

(24) Hughes, T. W.; Williamson, I. A. D.; Minkov, M.; Fan, S. Wave Physics as an Analog Recurrent Neural Network. *Science Advances* **2019**, *5* (12), eaay6946. https://doi.org/10.1126/sciadv.aay6946.

(25) Dong, J.; Rafayelyan, M.; Krzakala, F.; Gigan, S. Optical Reservoir Computing Using Multiple Light Scattering for Chaotic Systems Prediction. *IEEE Journal of Selected Topics in Quantum Electronics* **2020**, *26* (1), 1–12. https://doi.org/10.1109/JSTQE.2019.2936281.

(26) Teğin, U.; Yıldırım, M.; Oğuz, İ.; Moser, C.; Psaltis, D. Scalable Optical Learning Operator. *Nat Comput Sci* **2021**, *1* (8), 542–549. https://doi.org/10.1038/s43588-021-00112-0.

(27) Li, J.; Mengu, D.; Luo, Y.; Rivenson, Y.; Ozcan, A. Class-Specific Differential Detection in Diffractive Optical Neural Networks Improves Inference Accuracy. *AP* **2019**, *1* (4), 046001. https://doi.org/10.1117/1.AP.1.4.046001.

(28) Rahman, M. S. S.; Li, J.; Mengu, D.; Rivenson, Y.; Ozcan, A. Ensemble Learning of Diffractive Optical Networks. *Light Sci Appl* **2021**, *10* (1), 14. https://doi.org/10.1038/s41377-020-00446-w.

(29) Bai, B.; Li, Y.; Luo, Y.; Li, X.; Çetintaş, E.; Jarrahi, M.; Ozcan, A. All-Optical Image Classification through Unknown Random Diffusers Using a Single-Pixel Diffractive Network. *Light Sci Appl* **2023**, *12* (1), 69. https://doi.org/10.1038/s41377-023-01116-3.

(30) Li, J.; Mengu, D.; Yardimci, N. T.; Luo, Y.; Li, X.; Veli, M.; Rivenson, Y.; Jarrahi, M.; Ozcan, A. Spectrally Encoded Single-Pixel Machine Vision Using Diffractive Networks. *Science Advances* **2021**, *7* (13), eabd7690. https://doi.org/10.1126/sciadv.abd7690.

(31) Kulce, O.; Mengu, D.; Rivenson, Y.; Ozcan, A. All-Optical Information-Processing Capacity of Diffractive Surfaces. *Light Sci Appl* **2021**, *10* (1), 25. https://doi.org/10.1038/s41377-020-00439-9.

(32) Rahman, M. S. S.; Li, Y.; Yang, X.; Chen, S.; Ozcan, A. Massively Parallel and Universal Approximation of Nonlinear Functions Using Diffractive Processors. *eLight* **2025**, *5* (1), 32. https://doi.org/10.1186/s43593-025-00113-w.

(33) Shen, C.-Y.; Li, J.; Gan, T.; Li, Y.; Jarrahi, M.; Ozcan, A. All-Optical Phase Conjugation Using Diffractive Wavefront Processing. *Nat Commun* **2024**, *15* (1), 4989. https://doi.org/10.1038/s41467-024-49304-y.

(34) Mengu, D.; Ozcan, A. All-Optical Phase Recovery: Diffractive Computing for Quantitative Phase Imaging. *Advanced Optical Materials* **2022**, *10* (15), 2200281. https://doi.org/10.1002/adom.202200281.

(35) Shen, C.-Y.; Li, J.; Mengu, D.; Ozcan, A. Multispectral Quantitative Phase Imaging Using a Diffractive Optical Network. *Advanced Intelligent Systems* **2023**, *5* (11), 2300300. https://doi.org/10.1002/aisy.202300300.

(36) Li, Y.; Luo, Y.; Mengu, D.; Bai, B.; Ozcan, A. Quantitative Phase Imaging (QPI) through Random Diffusers Using a Diffractive Optical Network. *gxjzz* **2023**, *4* (3), 206–221. https://doi.org/10.37188/lam.2023.017.

(37) Işıl, Ç.; Mengu, D.; Zhao, Y.; Tabassum, A.; Li, J.; Luo, Y.; Jarrahi, M.; Ozcan, A. Super-Resolution Image Display Using Diffractive Decoders. *Science Advances* **2022**, *8* (48), eadd3433. https://doi.org/10.1126/sciadv.add3433.

(38) Bai, B.; Luo, Y.; Gan, T.; Hu, J.; Li, Y.; Zhao, Y.; Mengu, D.; Jarrahi, M.; Ozcan, A. To Image, or Not to Image: Class-Specific Diffractive Cameras with All-Optical Erasure of Undesired Objects. *eLight* **2022**, *2* (1), 14. https://doi.org/10.1186/s43593-022-00021-3.

(39) Bai, B.; Wei, H.; Yang, X.; Gan, T.; Mengu, D.; Jarrahi, M.; Ozcan, A. Data-Class-Specific All-Optical Transformations and Encryption. *Advanced Materials* **2023**, *35* (31), 2212091. https://doi.org/10.1002/adma.202212091.

(40) Gao, Y.; Jiao, S.; Fang, J.; Lei, T.; Xie, Z.; Yuan, X. Multiple-Image Encryption and Hiding with an Optical Diffractive Neural Network. *Optics Communications* **2020**, *463*, 125476. https://doi.org/10.1016/j.optcom.2020.125476.

(41) Luo, Y.; Mengu, D.; Ozcan, A. Cascadable All-Optical NAND Gates Using Diffractive Networks. *Sci Rep* **2022**, *12* (1), 7121. https://doi.org/10.1038/s41598-022-11331-4.

(42) Qian, C.; Lin, X.; Lin, X.; Xu, J.; Sun, Y.; Li, E.; Zhang, B.; Chen, H. Performing Optical Logic Operations by a Diffractive Neural Network. *Light Sci Appl* **2020**, *9* (1), 59. https://doi.org/10.1038/s41377-020-0303-2.

(43) Wang, P.; Xiong, W.; Huang, Z.; He, Y.; Xie, Z.; Liu, J.; Ye, H.; Li, Y.; Fan, D.; Chen, S. Orbital Angular Momentum Mode Logical Operation Using Optical Diffractive Neural Network. *Photon. Res., PRJ* **2021**, *9* (10), 2116–2124. https://doi.org/10.1364/PRJ.432919.

(44) Li, Y.; Li, J.; Zhao, Y.; Gan, T.; Hu, J.; Jarrahi, M.; Ozcan, A. Universal Polarization Transformations: Spatial Programming of Polarization Scattering Matrices Using a Deep Learning-Designed Diffractive Polarization Transformer. https://doi.org/10.1002/adma.202303395.

(45) Mengu, D.; Zhao, Y.; Yardimci, N. T.; Rivenson, Y.; Jarrahi, M.; Ozcan, A. Misalignment Resilient Diffractive Optical Networks. *Nanophotonics* **2020**, *9* (13), 4207–4219. https://doi.org/10.1515/nanoph-2020-0291.

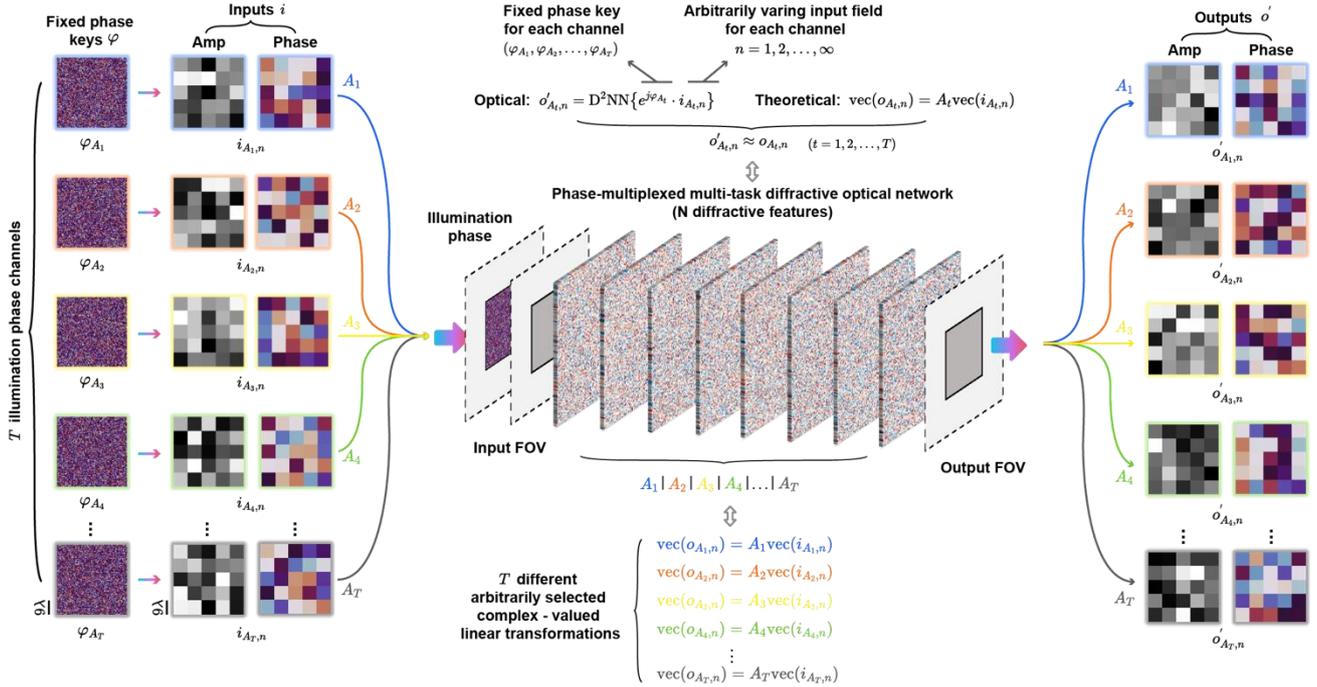

**Fig. 1 The schematic of a phase-multiplexed multi-task monochrome diffractive processor.** By encoding the illumination phase, the diffractive optical network reconfigures its output function and implements $T$ arbitrary complex-valued linear transformations between the input and output apertures using different illumination phase keys. The illumination phase keys, $\varphi_{A_1}, \varphi_{A_2}, \cdots, \varphi_{A_T}$, are jointly trained with the diffractive layers and then remain fixed; therefore, each phase key depends only on the linear transformation index $t$ and is sample-*independent*. In sequential operation, a plane-wave is modulated by one phase key $\varphi_{A_t}$ at a time to illuminate the input aperture, producing the channel-specific output field $o'_{A_t,n}$ at the diffractive network's output field of view (FOV) for a corresponding input $i_{A_t,n}$ field; here, $n = 1, 2, \dots, \infty$ enumerates the input samples/fields, indicating that the input samples can be chosen arbitrarily and there is no theoretical limit to the number of complex-valued input fields for which the optical transformations are accurate. By minimizing the error between the diffracted output field $o'_{A_t,n}$ and the target output $o_{A_t,n}$, the optically implemented mapping provides an accurate approximation to the target linear transformations $(A_1, A_2, \cdots, A_T)$.

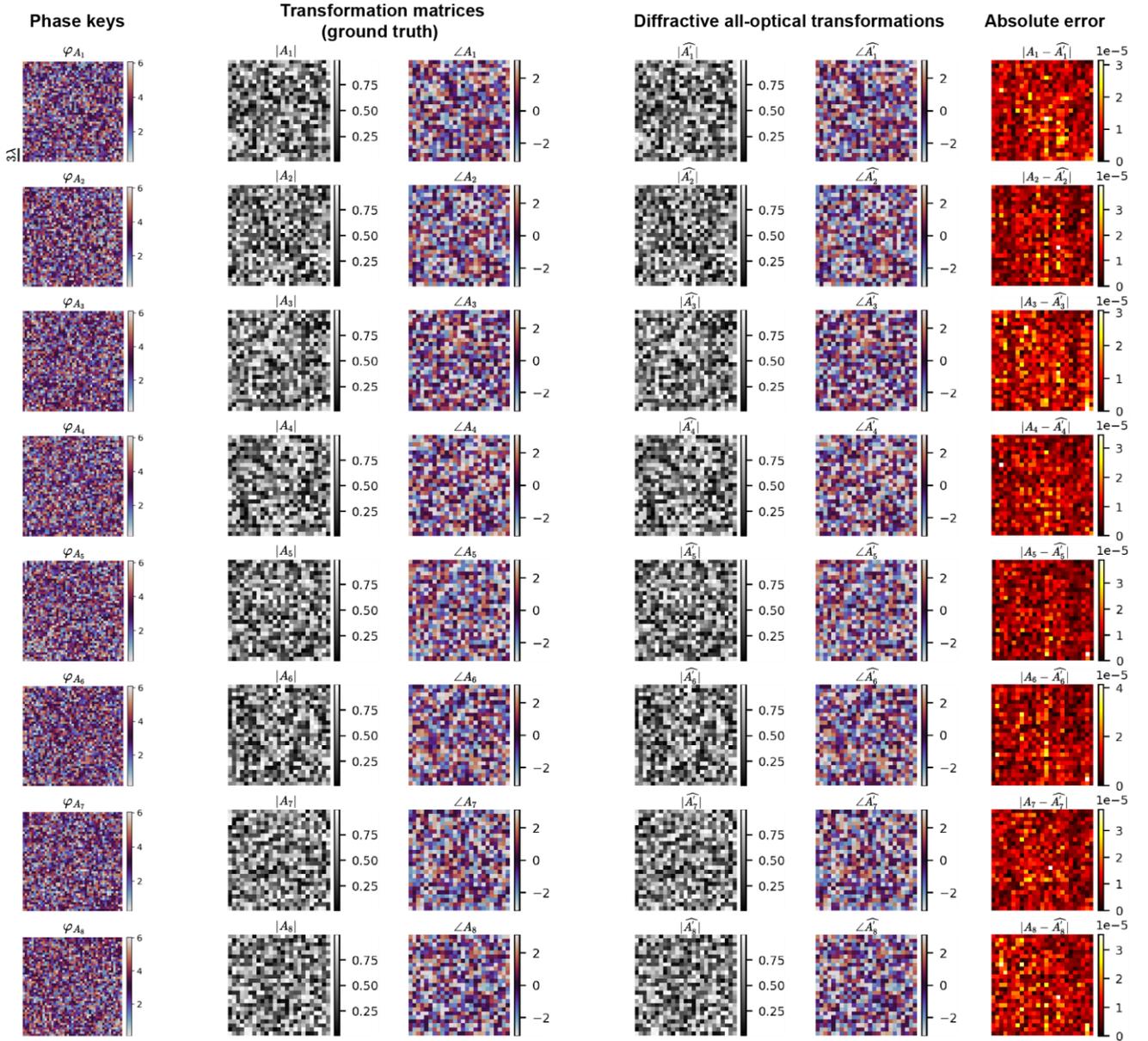

**Fig. 2 Complex-valued transformation matrices all-optically implemented by a phase-multiplexed monochrome diffractive processor with $T = 8$ and $N_i = N_o = 5^2$.** The diffractive network contains $N \approx 2TN_iN_o = 10,082$ trainable diffractive features. The first column shows the phase keys corresponding to the 8 transformation channels. Columns 4–5 present the complex-valued transformation matrices optically implemented by the diffractive network, while columns 2–3 show the corresponding ground-truth (target) matrices. Column 6 displays the absolute difference between each optically implemented matrix and its target matrix.

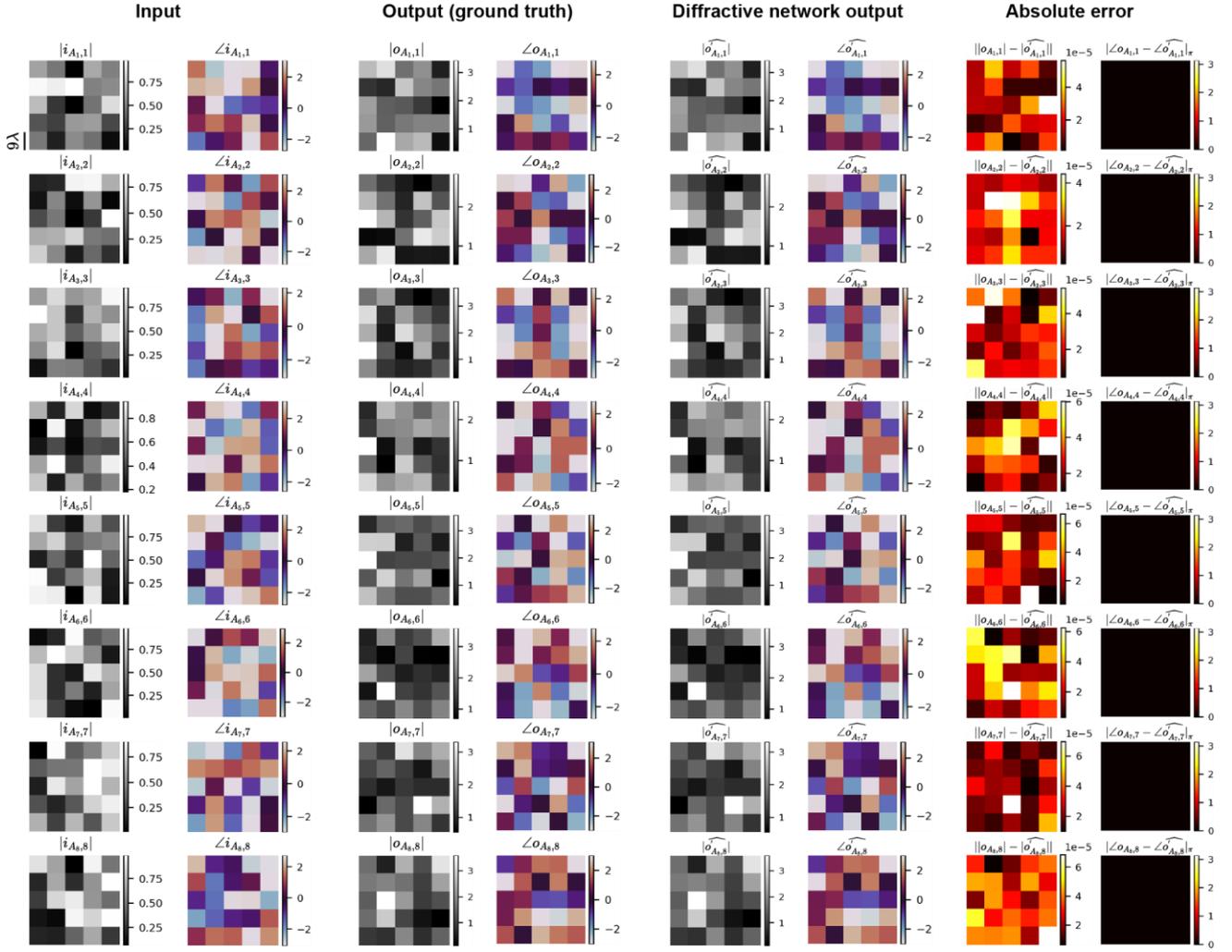

**Fig. 3 Examples of the input/output complex fields of the ground-truth (target) transformations, together with the all-optical output fields produced by the 8-channel phase-multiplexed monochrome diffractive design with $N \approx 2TN_iN_o = 10,082$ diffractive features.** Columns 7–8 show the absolute error between the ground-truth output fields and the all-optical outputs. The error level is negligible. $|\angle o - \angle \hat{o}'|_\pi$ denotes the wrapped phase difference between the ground-truth output field $o$ and the normalized diffractive-network output field $\hat{o}'$.

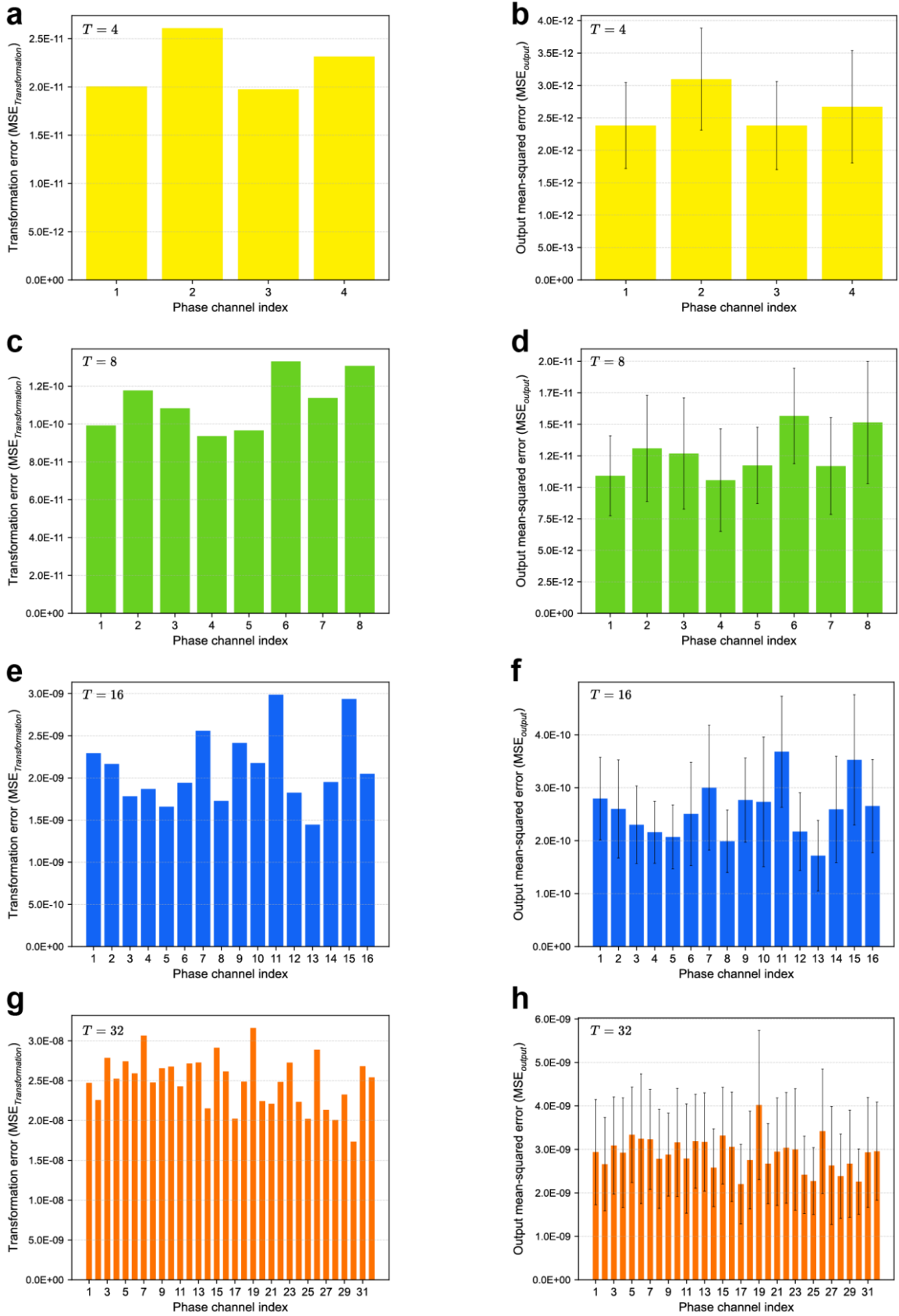

**Fig. 4 All-optical transformation performance of the monochrome diffractive network design with $N \approx 2TN_iN_o$ and $N_i = N_o = 5^2$ across different transformation channels.** For each phase-multiplexing configuration, the all-optical transformation error $\text{MSE}_{\text{Transformation}}$ and the output-field error $\text{MSE}_{\text{Output}}$ were computed. The standard

deviations (error bars) of these output-field errors are calculated over the entire blind test dataset. (a) and (b) correspond to $T = 4$, $N \approx 8N_iN_o$; (c) and (d) correspond to $T = 8$, $N \approx 16N_iN_o$; (e) and (f) correspond to $T = 16$, $N \approx 32N_iN_o$; and (g) and (h) correspond to $T = 32$, $N \approx 64N_iN_o$.

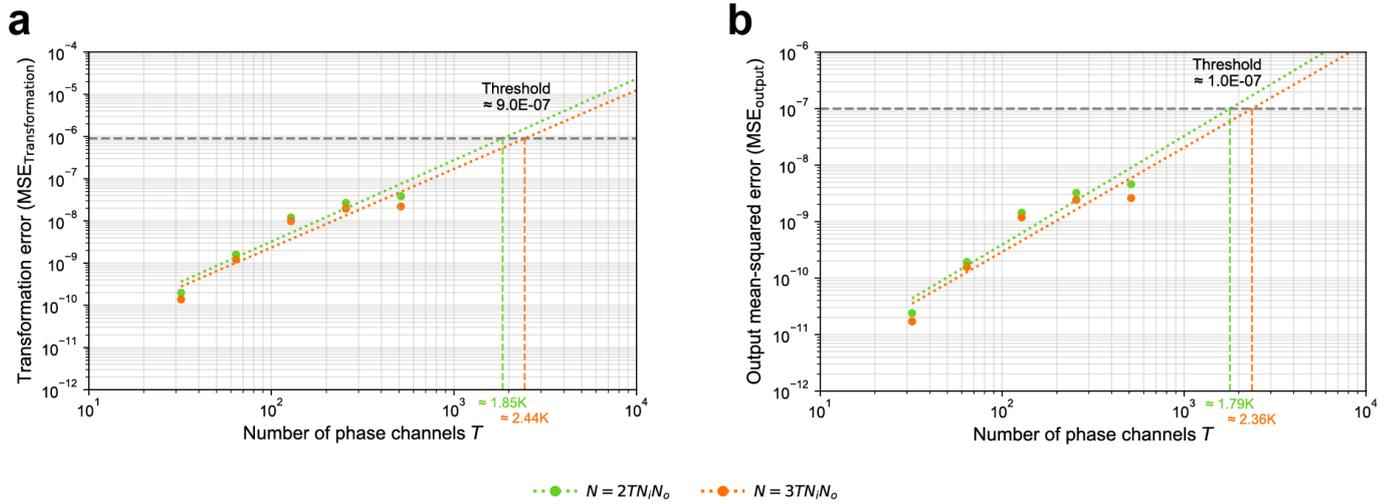

**Fig. 5 Exploration of the limits on the number of phase-multiplexed channels ($T$) achievable by a monochrome diffractive processor.** (a) The MSE between the ground-truth transformation matrices $\boldsymbol{A_t}$ and the all-optical transformations $\boldsymbol{A'_t}$ is reported as a function of the number of transformation channels $T \in \{32, 64, 128, 256, 512\}$. Results from two diffractive-network architectures with different numbers of diffractive features, $N \in \{2TN_iN_o, 3TN_iN_o\}$, are shown in green and orange, respectively. The corresponding dashed lines are fitted based on the data points for diffractive networks with $N = 2TN_iN_o$ and $N = 3TN_iN_o$. The intersection between the error threshold and the fitted lines indicates the potential upper limit of the number of transformation channels that the diffractive processor can support under a specific error bound. (b) Same as in (a), but reporting the MSE between the diffractive-network output fields and the ground-truth output fields.

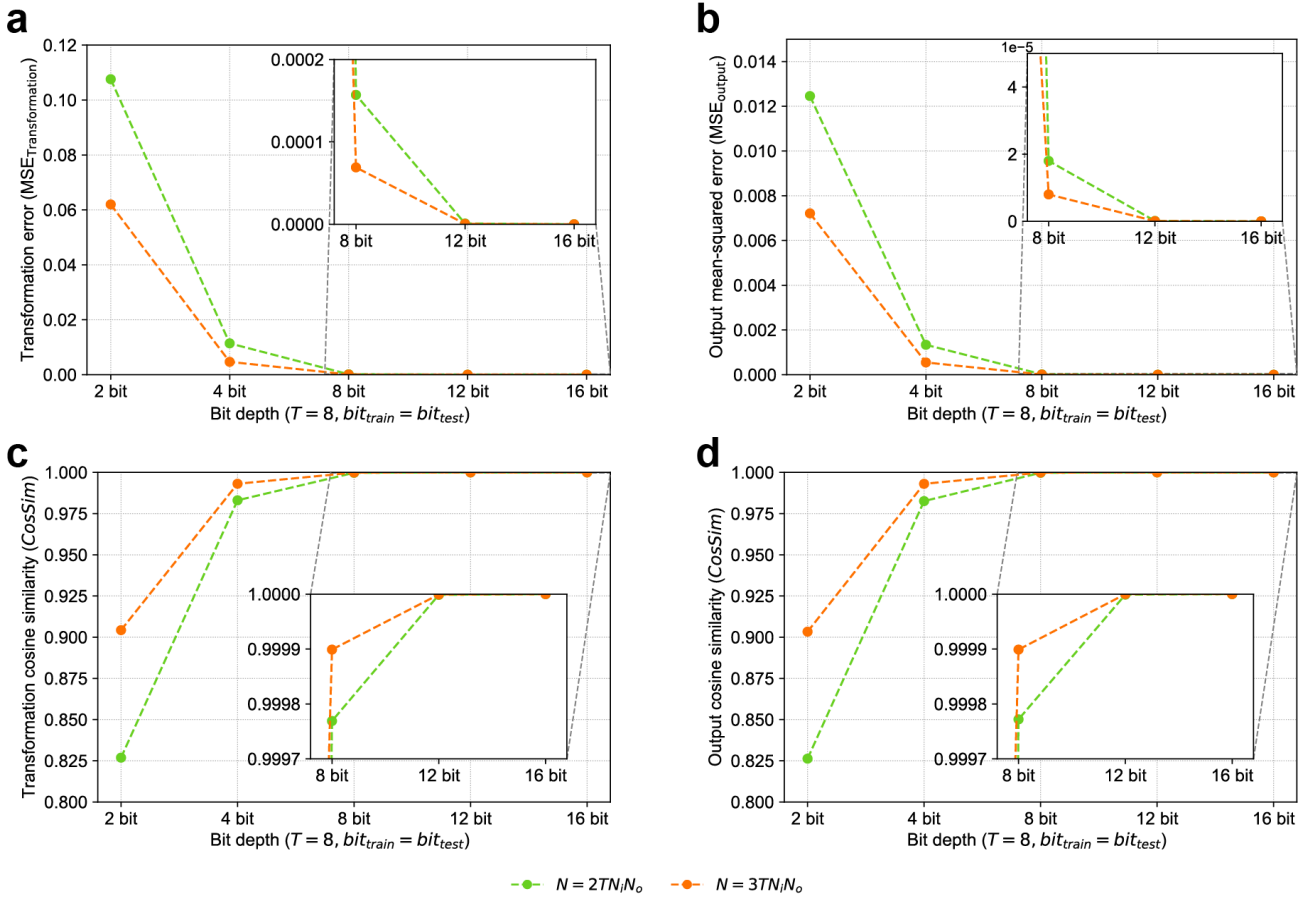

**Fig. 6 Effect of the layer bit depth on the all-optical linear transformation performance of phase-multiplexed monochrome diffractive processor designs ($T = 8, N \in \{2TN_iN_o, 3TN_iN_o\}$).** (a) The MSE between the ground-truth transformation matrices $A_t$ and the all-optical transformations $A'_t$ is reported as a function of the bit depth, $\{2, 4, 8, 12, 16\}$. The same bit depth is used during both the training and testing stages. A zoomed-in view of the curves is shown in the upper-right corner. (b) Same as (a), but the MSE values between the diffractive network output fields and the ground-truth output fields are reported. (c) and (d) are the corresponding cosine similarity (CosSim) values calculated for the transformation matrices and the output fields, respectively.

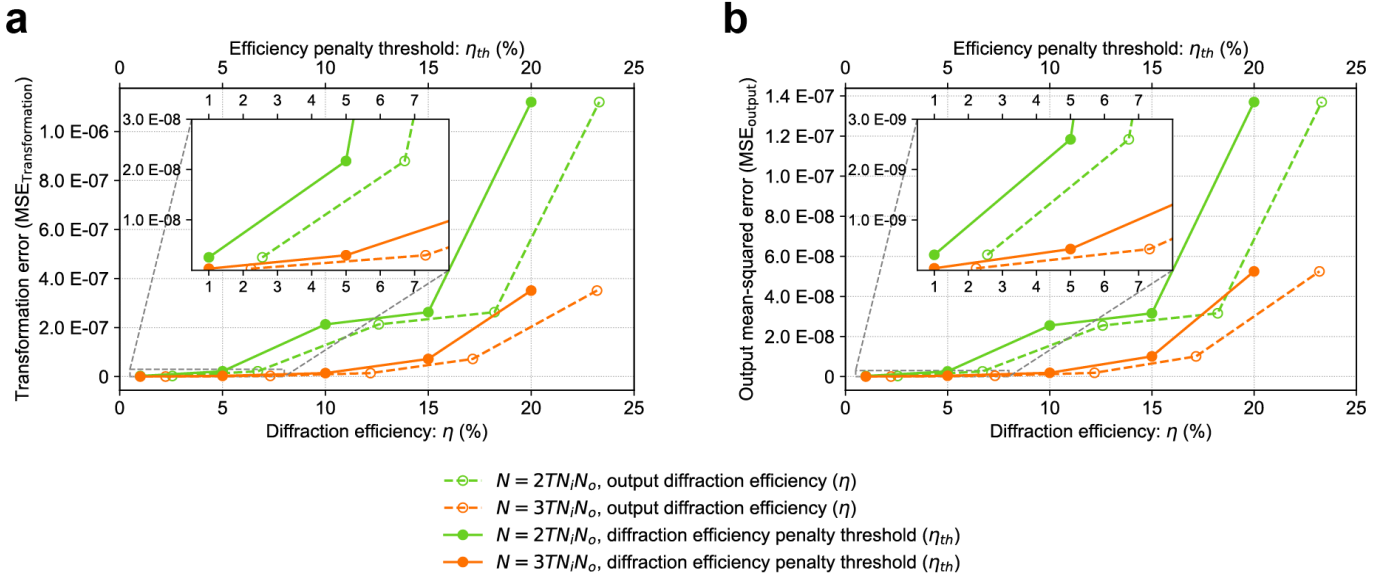

**Fig. 7 Trade-off between the output diffraction efficiency and the all-optical linear transformation accuracy of phase-multiplexed monochrome diffractive processor designs ($T = 8, N \in \{2TN_iN_o, 3TN_iN_o\}$).** (a) The MSE between the ground-truth transformation matrices $A_t$ and the all-optical transformations $A'_t$ is reported as a function of efficiency penalty threshold (solid lines, $\eta_{th} \in \{1\%, 5\%, 10\%, 15\%, 20\%\}$). The dashed lines represent the actual output diffraction efficiencies $\eta$ achieved by the diffractive networks (after their training) under the corresponding diffraction-efficiency penalty threshold, $\eta_{th}$. (b) Same as (a) but the MSE values between the diffractive network output fields and the ground-truth output fields are reported.

19